\documentclass[%
preprint,
 amsmath,amssymb,
 aps,
prb,
]{revtex4-1}
\usepackage{graphicx}
\usepackage{dcolumn}
\usepackage{bm}
\begin{document}
\preprint{APS/PRB/Weyl semimetals}

\title{ The large unsaturated magnetoresistance of Weyl semimetals }
\author{Yiming Pan}
\author{Huaiqiang Wang}
\author{Pengchao Lu}
\author{Jian Sun}
\author{Baigeng Wang}
\author{D. Y. Xing}
\affiliation{National Laboratory of Solid State Microstructures and Department of Physics,
Nanjing University, Nanjing 210093, China}

\date{\today}
\begin{abstract}
The Weyl semimetal (WSM) is a novel topological gapless state with  promises exotic transport due to chiral anomaly. Recently, a family of nonmagnetic WSM candidates including TaAs, NbAs, NbP etc is confirmed by first principle calculation and experiments. The TaAs family are reported to display the large unsaturated magnetoresistance (XMR), which have not yet been explained. Here we give a theoretical calculation of XMR based on the extended effective-medium approach to Weyl semimetals. We predict the power law of XMR at high magnetic field and the "turn-on" magnetic field, which are well in agreement with experiments data. Furthermore, we investigate the $\theta$-dependence magnetoresistance and find the transition between the postive XMR and the negative magnetoresistance induced by chiral anomaly, which should be confirmed by further experiments.
\begin{description}
\item[PACS numbers] 
71.20.-b,72.10.Bg,72.15.Gd,73.20.-r

\end{description}
\end{abstract}
\maketitle


Gapless topological phases, especially the Weyl semimetal(WSM), have attracted much interest\cite{YangBohmJung2014Classification, Hosur2013857}. Recently, TaAs-family materials have been proposed as possible candidates for realizing the WSM \cite{PhysRevX-5-011029, huang2015weyl}. TaAs is a kind of nonmagnetic material with preserved time reversal symmetry (TRS) and broken inversion symmetry (IS), and unlike previously proposed pyrochlore materials \cite{PhysRevB-83-205101}, its lattice structure satisfies a C4 rotation symmetry. According to the ab initio calculation\cite{PhysRevX-5-011029, huang2015weyl}, its Fermi surface is made up of 12 pairs of Weyl nodes, of which four pairs with tiny hole pockets (about 2 mev) are located near $\sigma$ point in the $k_z= 0$ plane and the other eight pairs with electron pockets are located symmetrically in the $k_z\neq 0$ plane. The exotic surface state of TaAs, the Fermi arc, has also been observed experimentally by ARPES\cite{xu2015discovery-sci, Xu2015Discovery-np, Lv2015Observation, Yang2015Weyl, PhysRevX.5.031013}. At the same time, it’s reported by several experiments on the transport property of TaAs, NbAs, NbP \cite{zhang2015tantalum, ghimire2015magnetotransport, Shekhar2015Extremely} that these candidate materials for WSM all show large unsaturated behavior of magneto-resistance (MR). To our knowledge, this near-linear behavior of MR in high magnetic field has not been theoretically explained up to now.

In Fig 1d, we show the experimental data of the MR of TaAs family materials from different groups \cite{zhang2015tantalum, ghimire2015magnetotransport, Shekhar2015Extremely} and for comparison, we also include the MR data of WTe2 \cite{Ali2014Large}. It’s easy to see that The MR of TaAs family is 1-2 orders of magnitude larger than that of WTe2. The unsaturated MR of single crystals in this family shows similar asymptotic behavior, namely, quadratic under low magnetic field ($<1T$) and near-linear under high field, with the transition point between 1T~10T. So far, there has been two kinds of explanations for the TaAs family’s unsaturated MR. One of them is based on the semi-classic two-band model with carrier compensation at n=p, and the charge resonance can explain the quadratic-type unsaturated behavior of MR quite well. Unfortunately, as the measurement of Hall conductivity at low temperatures shows electron-type doping, i.e. $n>p$, MR will get saturated in high field and the resonance condition is not satisfied here, so the carrier compensation theory is incapable of explaining the near-linear type MR. The other explanation stems from Abrikosov’s quantum MR model\cite{PhysRevB-58-2788}, which is based on linear dispersion with all charge carriers occupying the lowest Landau levels. This theory is characterised by the LMR’s stability against temperature, which, of course, cannot explain the temperature-dependent transition field H*. 

However, we find that the effective media theory (EMT) which was first developed in disorder systems could provide a simple physical explanation of the unsaturated MR of WSM \cite{PhysRevB-12-3368, PhysRevB-73-085202}. The current phase diagram for the asymptotic behavior of MR is shown in Fig.1e, including carrier compensation $H^2$( black dotted line)\cite{Ali2014Large, PhysRevLett-113-216601}, the Abrikosov LMR(red solid line)\cite{PhysRevB-58-2788}, the duality theorem for 2D systems(Olive dashed)\cite{PhysRevB-71-201304} and the classical saturated MR(blue short dashed)\cite{HuJ2008Classical}. Only the EMT can explain the asymptotic behavior of MR in the shaded area from saturated to quadratic type unsaturated. In addition, it can also give a quite reasonable estimate of the transition field H*. For WSM, when the magnetic field and electric filed are parallel, there also exists a negative MR induced by chiral anomaly, which has been observed in $Bi_{1-x}Sb_x$ materials~\cite{PhysRevLett-111-246603}. In this paper, we will generalize the EMT to systems with chiral anomaly and theoretically investigate the MR’s dependence on the angle between the magnetic field and electric field.

For Weyl semimetal, the total current includes two parts: the normal current $j_n$, and the anomalous current $j_a$. Thus total current j could be described by
\begin{equation}
j=j_n+j_a=\sigma E+\frac{e^2}{2\pi^2}b\times E+\frac{e^2 \Delta\mu}{2\pi^2}B
\end{equation}

Where the first term is normal current, sigma is the conductivity, the second term is the anomalous Hall effect, where b is the momentum seperation between two Weyl modes,and the last term is known as the chiral magnetic effect and $\Delta\mu$ is chiral energy shift. The last two terms are general anomalous current induced by chiral anomaly\cite{Hosur2013857}. The last term chiral magnetic effect is subtler and vanishes at equilibrium\cite{Hosur2013857, PhysRevB-89-035142}, but we could charging it with the external parallel magnetic field and electric field\cite{Hosur2013857}, and the chiral current $j_{ch}$ could be obtained by
\begin{equation}
j_{ch}=\gamma (E\cdot B)B
\end{equation}

Where  $\gamma=\frac{e^4 \tau_i}{4\pi^4 g(\epsilon_F)}, \tau_i$ is the intervalley-scattering time between two Weyl nodes, $g(\epsilon_F) $is the density of state at Fermi surface. The chiral energy difference is proportional to chiral anomaly $E\cdot B$ by charging, and the corresponding calculation are summarised in supplementary. And the total current then could be reduced in form $j=\sigma^{(t)} E$ , and the components of total conductivity are defined by
\begin{equation}
\sigma_{\alpha\beta}^{(t)}=\sigma_{\alpha\beta}-\varepsilon_{\alpha\beta\eta}b_{\eta}+\gamma B_{\alpha}B_{\beta}
\end{equation}

where $b_{\eta}$ is the redefined anomalous Hall conductivity (which is small than a quantum conductivity). In general, the total conductivity $\sigma^{(t)}$  does not satisfy Onsager reciprocal relations due to the second term violates the time-reversal symmetry. The first term normal conductivity could be obtained by the effective-medium approach,when the intra-node scattering $\tau$ is considering far greater than inter-node scattering $\tau_i$.

For Weyl semimetal candidates TaAs, eight pairs electron pockets and four pairs tiny hole pockets\cite{PhysRevX-5-011029} are simply considered as two bands model with the carrier concentration $n=\sum_{i=1} n_i, p=\sum_{i=1} p_i$ . Therefore, the electron (hole) band conductivity in magnetic field is given by
\begin{equation}
\sigma_{n}=\left(\begin{array}{ccc}\frac{\sigma_{0,n}}{1+\mu_n^2H^2} & \frac{\sigma_{0,n} \mu_n H}{1+\mu_n^2H^2} & 0 \\-\frac{\sigma_{0,n} \mu_n H}{1+\mu_n^2H^2} & \frac{\sigma_{0,n}}{1+\mu_n^2H^2} & 0 \\0 & 0 & \frac{\sigma_{0,n}}{1+\mu_n^2H^2}\end{array}\right),
\sigma_{p}=\left(\begin{array}{ccc}\frac{\sigma_{0,p}}{1+\mu_p^2H^2} & \frac{\sigma_{0,p} \mu_p H}{1+\mu_p^2H^2} & 0 \\-\frac{\sigma_{0,p} \mu_p H}{1+\mu_p^2H^2} & \frac{\sigma_{0,p}}{1+\mu_p^2H^2} & 0 \\0 & 0 & \frac{\sigma_{0,p}}{1+\mu_p^2H^2}\end{array}\right)
\end{equation}

Where $\mu_n,\mu_p$  are electron-pocket and hole-pocket mobility at zero magnetic field. The effective medium theory\cite{PhysRevB-12-3368} yields a self-consistent equation for effective normal conductivity,
\begin{equation}
\sum_{i=n,p}\delta\sigma_i (I-\Gamma \delta\sigma_i)=0
\end{equation}

Here, $\delta\sigma_i= \sigma_i-\sigma$  and $\Gamma$  is an ellipsoidal depolarization tensor which depend on the resulting conductivity. Thus, the total conductivity is combined the normal conductivity and the anomalous part in Eqs.3, then the magnetoresistance $MR=\frac{\rho_{xx}(H)-\rho_{xx}(0)}{\rho_{xx}(0)}$and the Hall coefficient $R_{H}=\frac{\rho_{xy}}{H}$, where $\rho(H)=(\sigma^{(t)})^{-1}$ .

First, we consider that H is perpendicular to E and then the total current have no the contribution of chiral anomaly. For WSM, the anomalous Hall conductivity is negligible due to $|b|<<\sigma_{xy}$, then the total current equals to the normal current approximately within the EMT. In Figure 2a, we assume that $\sigma_{0,n}=\sigma_{0,p},\mu=\mu_n=\mu_p$, then n=p. The magnetoresistance MR is as a function of magnetic field H at difference mobilities from $5\times10^3 cm^2V^{-1}s^{-1}$ to $10^6 cm^2V^{-1}s^{-1}$  in Figure 2a. We find that the power law of MR at fixed conpensation n=p tends to square dependence at high magnetic field, where the scaling $\alpha$  is 1.8 at high mobility $10^6 cm^2V^{-1}s^{-1}$ and 1.3 at low mobility $5\times10^3 cm^2V^{-1}s^{-1}$. Furthermore, at low magnetic field, the scaling $\alpha=2$  when the mobility $\mu>10^5 cm^2V^{-1}s^{-1}$ . Figure 2b shows that $R_H$ is alway zero at fixed n=p. The asymptotic behavior of MR confirm us the carrier compensation within semiclassic two band model\cite{Ali2014Large}. To further investigate the carrier compensation within EMT, we alter the resonance condition, and show in Figure 2c the MR as a function of hole mobility at different magnetic field 10T, 30T and 60T respectively and settle $\mu_n=10^5 cm^2V^{-1}s^{-1},\sigma_{0,n}=\sigma_{0,p}$. When $\mu_p$ reaches $10^5 cm^2V^{-1}s^{-1}$ ,corresponding to perfect compensation $p/n=1$ , the MR emerges a peak at high H due to charge resonance. When $\mu_p$ is away from $\mu_n$, then the MR is off-resonance and drops down quickly. Figure 2 shows that the square MR of carrier compensation within two-band model is the limit value of EMT at high mobility and fixed n=p.

Next, we apply the EMT to the magnetoresistance of WSM candidates. From the experimental estimation $n>p,\mu_n>\mu_p$ for TaAs family\cite{zhang2015tantalum}, the resonance condition is definitely unsatisfied. For simplicity, we normalized the electron-pocket conductivity $\sigma_{0,n}=en\mu_n=1$ and settle $\mu_n=10^5 cm^2V^{-1}s^{-1}$, then we study two cases: (i) at fixed $\mu_n=\mu_p$ , and varying $p/n$  ; (ii) at fixed p=n, and varying $\mu_n/\mu_p$ . We plot the calculating results with EMT in Figure 3. Figure 3a and 3b show the reduced resistivities $\rho_{xx}$ and $\rho_{xy}$ as a function of magnetic field at different ratios of carrier density $p/n$ . As $p/n$ decreasing, the scaling $\alpha$  at high magnetic field then decreases, correspondingly. Moreover, the power law of  $\rho_{xx}$ is still square ($\alpha'\approx 2$) at low magnetic field. The scaling changes from a parabolic to a less-linear dependence with a crossover manner characterized by a “turn-on” magnetic field H*. The H* and the scaling $\alpha$ at high magnetic field as a function of p/n are shown in Figure 3c. The “turn-on” magnetic varies roughly in the range from 1T to 10T, which are qualitatively consistent with the recant experimental data\cite{zhang2015tantalum} on TaAs, which the H* varies from 0.5T to 6T at different temperature. The corresponding power law at high magnetic field change from .5 to 1.5 without saturation. The quasi-linear power law is around at the ratio $p/n=0.9$. The Hall resistivity $\rho_{xy}$ in Figure 3b shows electron-type doping consistent with our assumption $n>p$.\\
For case (ii) with same carriers concentration n=p, Figure 3d and 3e show $\rho_{xx}$ and $\rho_{xy}$ as a function of magnetic field at different ratio of mobility $\mu_n=\mu_p$ . The resistivity $\rho_{xx}$  shows the power law $\alpha$ decrease with the mobility ratio, which are the similar asymptotic behavior in Figure 3a. The “turn-on” H* and scaling  as a function of the mobility ratio $\mu_n=\mu_p$  are shown in Figure 3f, which confirm the same power law with variation of p/n. The slight difference of   in Figure 3a and 3d is that the power law at low H decreases with the descending of mobility ratio, keeps no square any more. The Hall resistivity $\rho_{xy}$ in Figure 3e show hole-type doping characterization, even at fixed n=p.

Finally, we investigate the angle dependence of magnetoresistance of Weyl semimetals. When the magnetic field H is parallel to electric field E, the the total conductivity could reduced to $\sigma^{(t)}=\sigma_0+\gamma H^2$, where $\sigma_0$ is the conductivity at H=0. Therefore, the negative magnetoresistance induced by chiral anomaly is given by\cite{PhysRevLett-111-246603}
\begin{equation}
MR=-\frac{\gamma H^2}{\sigma_0+\gamma H^2}
\end{equation}

In general case for the angle $\theta$  between H and E, we consider the contribution of vertical magnetic field component $H\sin \theta$  to the normal conductivity part within the EMT, and the contribution of the parallel component $H\cos \theta$ to the anomalous conductivity\cite{liang2015ultrahigh}. In Figure 4a, we show the chiral energy difference $\Delta\mu$  between two Weyl nodes could be charged by chiral anomaly. Figure 4b show the magnetoresistance with normalized $\sigma_{n,p}=1, \mu_{n,p}=10^5 cm^2V^{-1}s^{-1}$  and settle the parameter $\gamma=10^{-4}$ at different angle $\theta=0$  to $\pi/2$. The points show the angle dependence of MR with chiral anomaly, while the dashed lines are referenced MR without the contribution of chiral anomaly. At small angle the MR shows a transition from positive MR to negative MR at low magnetic field. For the conditions of carrier compensation, we could give an explicit expression for the total conductivity $\sigma^{(t)}=\frac{\sigma_0}{1+\mu^2H^2 \sin^2 \theta}+\gamma H^2\cos^2 \theta$, then the $\theta-$dependence magnetoresistance
\begin{equation}
MR=\frac{1}{\frac{1}{1+\mu^2 H^2 \sin^2 \theta}+\mu'^2H^2\cos^2 \theta}-1
\end{equation}

Where $\mu'=\sqrt{\gamma/\sigma_0}$ is the effective mobility. The magnetoresistance transition could be obtained by $\frac{\partial MR}{\partial H}=0$ . Figure 4c shows the contour plot of MR transition between PMR and NMR, and the yellow solid line shows the transition position as a function of angle $\theta$ and magnetic field H with dimensionless parameter $\delta=\mu'/\mu=0.3$. Figure 4d show the transition curves at different parameter $\delta$ from 0.1 to 1.0. The contribution of chiral anomaly increases with $\delta$ increasing, and the magnetoresistance show more NMR characterization. We need solve the EMT self-consistent equations for the general cases when the carrier compensation condition is unsatisfied (i.e. $n>p,\mu_n>\mu_p$). For experimental realization on TaAs family, however, the anomaly-induced negative magnetoresistance has to be carefully measured because it is much smaller than the giant unsaturated PMR as a angle-dependence ‘background’. The parameter $\delta\ll 1$  shows that the negative MR is not easy to detected in experiments.

In conclusion, we extended the effective medium theory into Weyl semimetals with chiral anomaly. Our theory could give the large unsaturated magnetoresistance of Weyl semimetals, like TaAs, NbAs, NbP etc. The power law at high magnetic field and “turn-on” magnetic field are studied and in qualitatively agreement with experimental data. Furthermore, we investigated the angle-dependence MR. Theoretically, there is a transition from postive MR to negative MR when the chiral anomaly is not negligible to compare with the large positive magnetoresistance ‘background’. The extended EMT could give a good explanation to magnetotransport of current experiments on Weyl semimetals candidates TaAs, NbAs, NbP etc.

We thank Q.Q. Cheng for drawing figures very much. This work was supported by 973 Program (Grant No. 2011CB922103),  the National Natural Science Foundation of China (Grant Nos. 60825402, 11023002, 51372112 and 11574133), the National Key Projects for Basic Research in China (Grant No.2015CB921202), the NSF of Jiangsu Province (Grant No. BK20150012), and the Fundamental Research Funds for the Central Universities for financially supporting.




\bibliography{apssamp2222}

\begin{thebibliography}{23}%
\makeatletter
\providecommand \@ifxundefined [1]{%
 \@ifx{#1\undefined}
}%
\providecommand \@ifnum [1]{%
 \ifnum #1\expandafter \@firstoftwo
 \else \expandafter \@secondoftwo
 \fi
}%
\providecommand \@ifx [1]{%
 \ifx #1\expandafter \@firstoftwo
 \else \expandafter \@secondoftwo
 \fi
}%
\providecommand \natexlab [1]{#1}%
\providecommand \enquote  [1]{``#1''}%
\providecommand \bibnamefont  [1]{#1}%
\providecommand \bibfnamefont [1]{#1}%
\providecommand \citenamefont [1]{#1}%
\providecommand \href@noop [0]{\@secondoftwo}%
\providecommand \href [0]{\begingroup \@sanitize@url \@href}%
\providecommand \@href[1]{\@@startlink{#1}\@@href}%
\providecommand \@@href[1]{\endgroup#1\@@endlink}%
\providecommand \@sanitize@url [0]{\catcode `\\12\catcode `\$12\catcode
  `\&12\catcode `\#12\catcode `\^12\catcode `\_12\catcode `\%12\relax}%
\providecommand \@@startlink[1]{}%
\providecommand \@@endlink[0]{}%
\providecommand \url  [0]{\begingroup\@sanitize@url \@url }%
\providecommand \@url [1]{\endgroup\@href {#1}{\urlprefix }}%
\providecommand \urlprefix  [0]{URL }%
\providecommand \Eprint [0]{\href }%
\providecommand \doibase [0]{http://dx.doi.org/}%
\providecommand \selectlanguage [0]{\@gobble}%
\providecommand \bibinfo  [0]{\@secondoftwo}%
\providecommand \bibfield  [0]{\@secondoftwo}%
\providecommand \translation [1]{[#1]}%
\providecommand \BibitemOpen [0]{}%
\providecommand \bibitemStop [0]{}%
\providecommand \bibitemNoStop [0]{.\EOS\space}%
\providecommand \EOS [0]{\spacefactor3000\relax}%
\providecommand \BibitemShut  [1]{\csname bibitem#1\endcsname}%
\let\auto@bib@innerbib\@empty
\bibitem [{\citenamefont {Yang}\ and\ \citenamefont
  {et~al}(2014)}]{YangBohmJung2014Classification}%
  \BibitemOpen
  \bibfield  {author} {\bibinfo {author} {\bibnamefont {Yang}}\ and\ \bibinfo
  {author} {\bibnamefont {et~al}},\ }\href@noop {} {\bibfield  {journal}
  {\bibinfo  {journal} {Nat. Commun.}\ }\textbf {\bibinfo {volume} {5}}
  (\bibinfo {year} {2014})}\BibitemShut {NoStop}%
\bibitem [{\citenamefont {Hosur}\ and\ \citenamefont
  {Qi}(2013)}]{Hosur2013857}%
  \BibitemOpen
  \bibfield  {author} {\bibinfo {author} {\bibfnamefont {P.}~\bibnamefont
  {Hosur}}\ and\ \bibinfo {author} {\bibfnamefont {X.}~\bibnamefont {Qi}},\
  }\href@noop {} {\bibfield  {journal} {\bibinfo  {journal} {Comptes Rendus
  Physique}\ }\textbf {\bibinfo {volume} {14}},\ \bibinfo {pages} {857}
  (\bibinfo {year} {2013})}\BibitemShut {NoStop}%
\bibitem [{\citenamefont {Weng}\ and\ \citenamefont
  {et~al}(2015)}]{PhysRevX-5-011029}%
  \BibitemOpen
  \bibfield  {author} {\bibinfo {author} {\bibfnamefont {H.}~\bibnamefont
  {Weng}}\ and\ \bibinfo {author} {\bibnamefont {et~al}},\ }\href {\doibase
  10.1103/PhysRevX.5.011029} {\bibfield  {journal} {\bibinfo  {journal} {Phys.
  Rev. X}\ }\textbf {\bibinfo {volume} {5}},\ \bibinfo {pages} {011029}
  (\bibinfo {year} {2015})}\BibitemShut {NoStop}%
\bibitem [{\citenamefont {Huang}\ and\ \citenamefont
  {et~al}(2015)}]{huang2015weyl}%
  \BibitemOpen
  \bibfield  {author} {\bibinfo {author} {\bibfnamefont {S.-M.}\ \bibnamefont
  {Huang}}\ and\ \bibinfo {author} {\bibnamefont {et~al}},\ }\href@noop {}
  {\bibfield  {journal} {\bibinfo  {journal} {Nat. Commun.}\ }\textbf {\bibinfo
  {volume} {6}} (\bibinfo {year} {2015})}\BibitemShut {NoStop}%
\bibitem [{\citenamefont {Wan}\ and\ \citenamefont
  {et~al}(2011)}]{PhysRevB-83-205101}%
  \BibitemOpen
  \bibfield  {author} {\bibinfo {author} {\bibfnamefont {X.}~\bibnamefont
  {Wan}}\ and\ \bibinfo {author} {\bibnamefont {et~al}},\ }\href {\doibase
  10.1103/PhysRevB.83.205101} {\bibfield  {journal} {\bibinfo  {journal} {Phys.
  Rev. B}\ }\textbf {\bibinfo {volume} {83}},\ \bibinfo {pages} {205101}
  (\bibinfo {year} {2011})}\BibitemShut {NoStop}%
\bibitem [{\citenamefont {Xu}\ and\ \citenamefont
  {et~al}(2015{\natexlab{a}})}]{xu2015discovery-sci}%
  \BibitemOpen
  \bibfield  {author} {\bibinfo {author} {\bibfnamefont {S.-Y.}\ \bibnamefont
  {Xu}}\ and\ \bibinfo {author} {\bibnamefont {et~al}},\ }\href@noop {}
  {\bibfield  {journal} {\bibinfo  {journal} {Science}\ }\textbf {\bibinfo
  {volume} {349}},\ \bibinfo {pages} {613} (\bibinfo {year}
  {2015}{\natexlab{a}})}\BibitemShut {NoStop}%
\bibitem [{\citenamefont {Xu}\ and\ \citenamefont
  {et~al}(2015{\natexlab{b}})}]{Xu2015Discovery-np}%
  \BibitemOpen
  \bibfield  {author} {\bibinfo {author} {\bibfnamefont {S.-Y.}\ \bibnamefont
  {Xu}}\ and\ \bibinfo {author} {\bibnamefont {et~al}},\ }\href
  {http://dx.doi.org/10.1038/nphys3437} {\bibfield  {journal} {\bibinfo
  {journal} {Nat. Phys.}\ }\textbf {\bibinfo {volume} {11}},\ \bibinfo {pages}
  {748} (\bibinfo {year} {2015}{\natexlab{b}})}\BibitemShut {NoStop}%
\bibitem [{\citenamefont {Lv}\ and\ \citenamefont
  {et~al}(2015{\natexlab{a}})}]{Lv2015Observation}%
  \BibitemOpen
  \bibfield  {author} {\bibinfo {author} {\bibfnamefont {B.~Q.}\ \bibnamefont
  {Lv}}\ and\ \bibinfo {author} {\bibnamefont {et~al}},\ }\href
  {http://dx.doi.org/10.1038/nphys3426} {\bibfield  {journal} {\bibinfo
  {journal} {Nat. Phys.}\ }\textbf {\bibinfo {volume} {11}},\ \bibinfo {pages}
  {724} (\bibinfo {year} {2015}{\natexlab{a}})}\BibitemShut {NoStop}%
\bibitem [{\citenamefont {Yang}\ and\ \citenamefont
  {et~al}(2015)}]{Yang2015Weyl}%
  \BibitemOpen
  \bibfield  {author} {\bibinfo {author} {\bibfnamefont {L.~X.}\ \bibnamefont
  {Yang}}\ and\ \bibinfo {author} {\bibnamefont {et~al}},\ }\href
  {http://dx.doi.org/10.1038/nphys3425} {\bibfield  {journal} {\bibinfo
  {journal} {Nat. Phys.}\ }\textbf {\bibinfo {volume} {11}},\ \bibinfo {pages}
  {728} (\bibinfo {year} {2015})}\BibitemShut {NoStop}%
\bibitem [{\citenamefont {Lv}\ and\ \citenamefont
  {et~al}(2015{\natexlab{b}})}]{PhysRevX.5.031013}%
  \BibitemOpen
  \bibfield  {author} {\bibinfo {author} {\bibfnamefont {B.~Q.}\ \bibnamefont
  {Lv}}\ and\ \bibinfo {author} {\bibnamefont {et~al}},\ }\href {\doibase
  10.1103/PhysRevX.5.031013} {\bibfield  {journal} {\bibinfo  {journal} {Phys.
  Rev. X}\ }\textbf {\bibinfo {volume} {5}},\ \bibinfo {pages} {031013}
  (\bibinfo {year} {2015}{\natexlab{b}})}\BibitemShut {NoStop}%
\bibitem [{\citenamefont {Zhang}\ and\ \citenamefont
  {et~al}(2015)}]{zhang2015tantalum}%
  \BibitemOpen
  \bibfield  {author} {\bibinfo {author} {\bibfnamefont {C.}~\bibnamefont
  {Zhang}}\ and\ \bibinfo {author} {\bibnamefont {et~al}},\ }\href@noop {}
  {\bibfield  {journal} {\bibinfo  {journal} {arXiv:1502.00251}\ } (\bibinfo
  {year} {2015})}\BibitemShut {NoStop}%
\bibitem [{\citenamefont {Ghimire}\ and\ \citenamefont
  {et~al}(2015)}]{ghimire2015magnetotransport}%
  \BibitemOpen
  \bibfield  {author} {\bibinfo {author} {\bibfnamefont {N.}~\bibnamefont
  {Ghimire}}\ and\ \bibinfo {author} {\bibnamefont {et~al}},\ }\href@noop {}
  {\bibfield  {journal} {\bibinfo  {journal} {Journal of Physics: Condensed
  Matter}\ }\textbf {\bibinfo {volume} {27}},\ \bibinfo {pages} {152201}
  (\bibinfo {year} {2015})}\BibitemShut {NoStop}%
\bibitem [{\citenamefont {Shekhar}\ and\ \citenamefont
  {et~al}(2015)}]{Shekhar2015Extremely}%
  \BibitemOpen
  \bibfield  {author} {\bibinfo {author} {\bibfnamefont {C.}~\bibnamefont
  {Shekhar}}\ and\ \bibinfo {author} {\bibnamefont {et~al}},\ }\href
  {http://dx.doi.org/10.1038/nphys3372} {\bibfield  {journal} {\bibinfo
  {journal} {Nat. Phys.}\ }\textbf {\bibinfo {volume} {11}},\ \bibinfo {pages}
  {645} (\bibinfo {year} {2015})}\BibitemShut {NoStop}%
\bibitem [{\citenamefont {Ali}\ and\ \citenamefont
  {et~al}(2014)}]{Ali2014Large}%
  \BibitemOpen
  \bibfield  {author} {\bibinfo {author} {\bibfnamefont {M.~N.}\ \bibnamefont
  {Ali}}\ and\ \bibinfo {author} {\bibnamefont {et~al}},\ }\href@noop {}
  {\bibfield  {journal} {\bibinfo  {journal} {Nature}\ }\textbf {\bibinfo
  {volume} {514}},\ \bibinfo {pages} {205} (\bibinfo {year}
  {2014})}\BibitemShut {NoStop}%
\bibitem [{\citenamefont {Abrikosov}(1998)}]{PhysRevB-58-2788}%
  \BibitemOpen
  \bibfield  {author} {\bibinfo {author} {\bibfnamefont {A.~A.}\ \bibnamefont
  {Abrikosov}},\ }\href {\doibase 10.1103/PhysRevB.58.2788} {\bibfield
  {journal} {\bibinfo  {journal} {Phys. Rev. B}\ }\textbf {\bibinfo {volume}
  {58}},\ \bibinfo {pages} {2788} (\bibinfo {year} {1998})}\BibitemShut
  {NoStop}%
\bibitem [{\citenamefont {Stroud}(1975)}]{PhysRevB-12-3368}%
  \BibitemOpen
  \bibfield  {author} {\bibinfo {author} {\bibfnamefont {D.}~\bibnamefont
  {Stroud}},\ }\href {\doibase 10.1103/PhysRevB.12.3368} {\bibfield  {journal}
  {\bibinfo  {journal} {Phys. Rev. B}\ }\textbf {\bibinfo {volume} {12}},\
  \bibinfo {pages} {3368} (\bibinfo {year} {1975})}\BibitemShut {NoStop}%
\bibitem [{\citenamefont {Guttal}\ and\ \citenamefont
  {Stroud}(2006)}]{PhysRevB-73-085202}%
  \BibitemOpen
  \bibfield  {author} {\bibinfo {author} {\bibfnamefont {V.}~\bibnamefont
  {Guttal}}\ and\ \bibinfo {author} {\bibfnamefont {D.}~\bibnamefont
  {Stroud}},\ }\href {\doibase 10.1103/PhysRevB.73.085202} {\bibfield
  {journal} {\bibinfo  {journal} {Phys. Rev. B}\ }\textbf {\bibinfo {volume}
  {73}},\ \bibinfo {pages} {085202} (\bibinfo {year} {2006})}\BibitemShut
  {NoStop}%
\bibitem [{\citenamefont {Pletikosi{\'c}}\ and\ \citenamefont
  {et~al}(2014)}]{PhysRevLett-113-216601}%
  \BibitemOpen
  \bibfield  {author} {\bibinfo {author} {\bibfnamefont {I.}~\bibnamefont
  {Pletikosi{\'c}}}\ and\ \bibinfo {author} {\bibnamefont {et~al}},\
  }\href@noop {} {\bibfield  {journal} {\bibinfo  {journal} {Phys. Rev. Lett.}\
  }\textbf {\bibinfo {volume} {113}},\ \bibinfo {pages} {216601} (\bibinfo
  {year} {2014})}\BibitemShut {NoStop}%
\bibitem [{\citenamefont {Guttal}\ and\ \citenamefont
  {Stroud}(2005)}]{PhysRevB-71-201304}%
  \BibitemOpen
  \bibfield  {author} {\bibinfo {author} {\bibfnamefont {V.}~\bibnamefont
  {Guttal}}\ and\ \bibinfo {author} {\bibfnamefont {D.}~\bibnamefont
  {Stroud}},\ }\href {\doibase 10.1103/PhysRevB.71.201304} {\bibfield
  {journal} {\bibinfo  {journal} {Phys. Rev. B}\ }\textbf {\bibinfo {volume}
  {71}},\ \bibinfo {pages} {201304} (\bibinfo {year} {2005})}\BibitemShut
  {NoStop}%
\bibitem [{\citenamefont {Hu}\ and\ \citenamefont
  {Rosenbaum}(2008)}]{HuJ2008Classical}%
  \BibitemOpen
  \bibfield  {author} {\bibinfo {author} {\bibfnamefont {J.}~\bibnamefont
  {Hu}}\ and\ \bibinfo {author} {\bibfnamefont {T.}~\bibnamefont {Rosenbaum}},\
  }\href@noop {} {\bibfield  {journal} {\bibinfo  {journal} {Nat. Mater.}\
  }\textbf {\bibinfo {volume} {7}},\ \bibinfo {pages} {697} (\bibinfo {year}
  {2008})}\BibitemShut {NoStop}%
\bibitem [{\citenamefont {Kim}\ and\ \citenamefont
  {et~al}(2013)}]{PhysRevLett-111-246603}%
  \BibitemOpen
  \bibfield  {author} {\bibinfo {author} {\bibfnamefont {H.-J.}\ \bibnamefont
  {Kim}}\ and\ \bibinfo {author} {\bibnamefont {et~al}},\ }\href {\doibase
  10.1103/PhysRevLett.111.246603} {\bibfield  {journal} {\bibinfo  {journal}
  {Phys. Rev. Lett.}\ }\textbf {\bibinfo {volume} {111}},\ \bibinfo {pages}
  {246603} (\bibinfo {year} {2013})}\BibitemShut {NoStop}%
\bibitem [{\citenamefont {Ba{\c{s}}ar}\ and\ \citenamefont
  {et~al}(2014)}]{PhysRevB-89-035142}%
  \BibitemOpen
  \bibfield  {author} {\bibinfo {author} {\bibfnamefont {G.}~\bibnamefont
  {Ba{\c{s}}ar}}\ and\ \bibinfo {author} {\bibnamefont {et~al}},\ }\href@noop
  {} {\bibfield  {journal} {\bibinfo  {journal} {Phys. Rev. B}\ }\textbf
  {\bibinfo {volume} {89}},\ \bibinfo {pages} {035142} (\bibinfo {year}
  {2014})}\BibitemShut {NoStop}%
\bibitem [{\citenamefont {Liang}\ and\ \citenamefont
  {et~al}(2015)}]{liang2015ultrahigh}%
  \BibitemOpen
  \bibfield  {author} {\bibinfo {author} {\bibfnamefont {T.}~\bibnamefont
  {Liang}}\ and\ \bibinfo {author} {\bibnamefont {et~al}},\ }\href@noop {}
  {\bibfield  {journal} {\bibinfo  {journal} {Nat. Mater.}\ }\textbf {\bibinfo
  {volume} {14}},\ \bibinfo {pages} {280} (\bibinfo {year} {2015})}\BibitemShut
  {NoStop}%
\end{thebibliography}%

\newpage
\begin{figure}[t]
\begin{center}\includegraphics[width=\columnwidth]{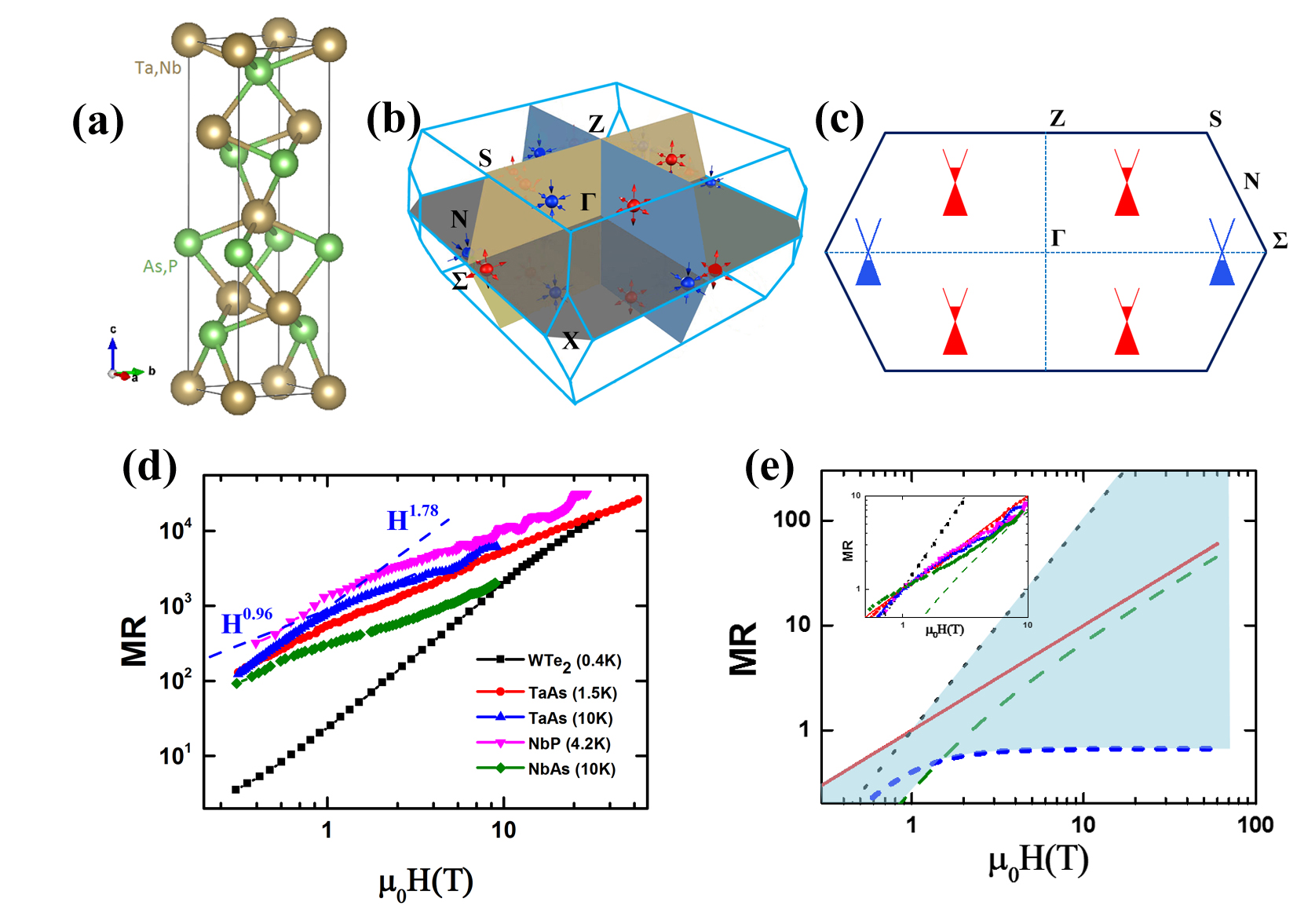}
\end{center}
\caption{Structure and magnetoresistance of TaAs family. (a)Body-centred tetragonal structure of Ta(Nb)As(P). Ta(Nb) and As(P) atoms are represented by brown and green atoms. (b) The first Brillouin zone (BZ) in the momentum space reprinted from Ref.~\cite{PhysRevX-5-011029}. Red and blue dots stand for Weyl nodes with opposite chirality. (c) Side view of BZ for Fermi surface. The Weyl nodes in the kz=0 plane form hole pockets, while the others form electron pockets symmetrically.(d)Experimental data on magnetoresistance of TaAs faimly at low temperature, TaAs (red,blue) from Ref.~\cite{zhang2015tantalum}, NbP(magenta) from Ref.~\cite{Shekhar2015Extremely}, NbAs(Olivia) from Ref.~\cite{ghimire2015magnetotransport}, and $\text{WTe}_2$(black) from Ref.~\cite{Ali2014Large}. (d)Power law of the unsaturated magnetoresistance, the square dependence from carrier compensation(black dotted), the exactly linear dependence from Abrikosov’s quantum limit (red solid), the quasi-linear dependence from duality theorem on 2D electron systems (Olivia dashed), and saturated MR of classical prediction (blue short-dashed). The shadow region are well predicted by the effective-medium theory. Inset shows the noramlized MR of TaAs family are quasi-linear above at 1T, while WTe2 are perfectly square dependence.}
\label{fig:fig1}
\end{figure}

\begin{figure}[t]
\begin{center}\includegraphics[width=\columnwidth]{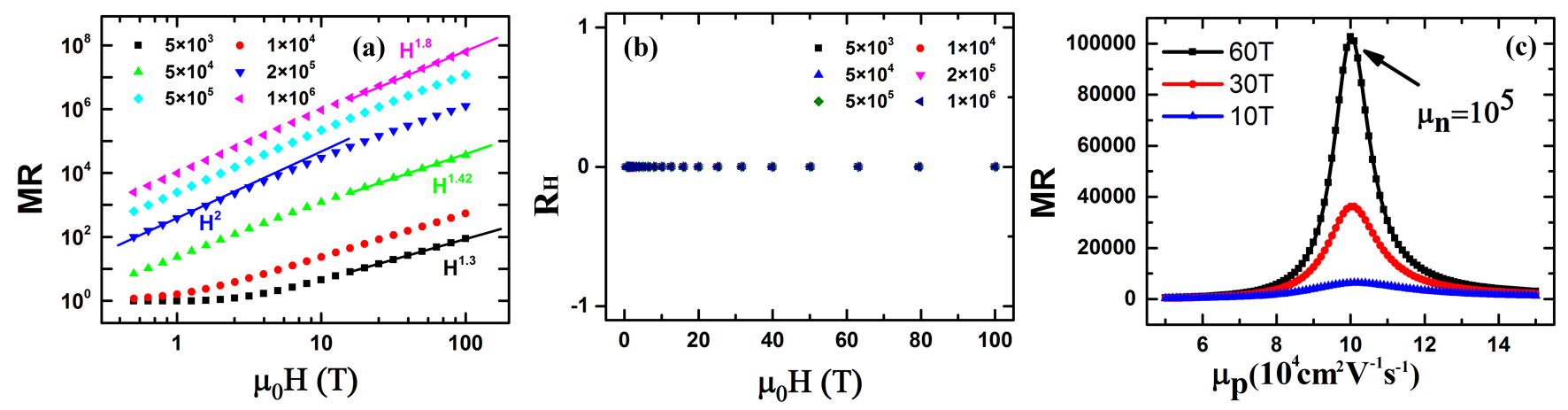}
\end{center}
\caption{The magnetoresistance (MR) on perfect n-p compensation condition. (a) MR as a function of magnetic field at different mobilities from $5\times10^3cm^{2}V^{-1}s^{-1}$ to $10^6cm^{2}V^{-1}s^{-1}$. (b) The corresponding Hall resistance $\rho_xy$ are all zeros at fixed n=p.(c)The electron conductivity and hole conductivity are equal, and the electron mobility is fixed at $10^5cm^{2}V^{-1}s^{-1}$. MR as a function of hole mobility at different magnetic field 10T, 30T and 60T.}
\label{fig:fig2}
\end{figure}

\begin{figure}[t]
\begin{center}\includegraphics[width=\columnwidth]{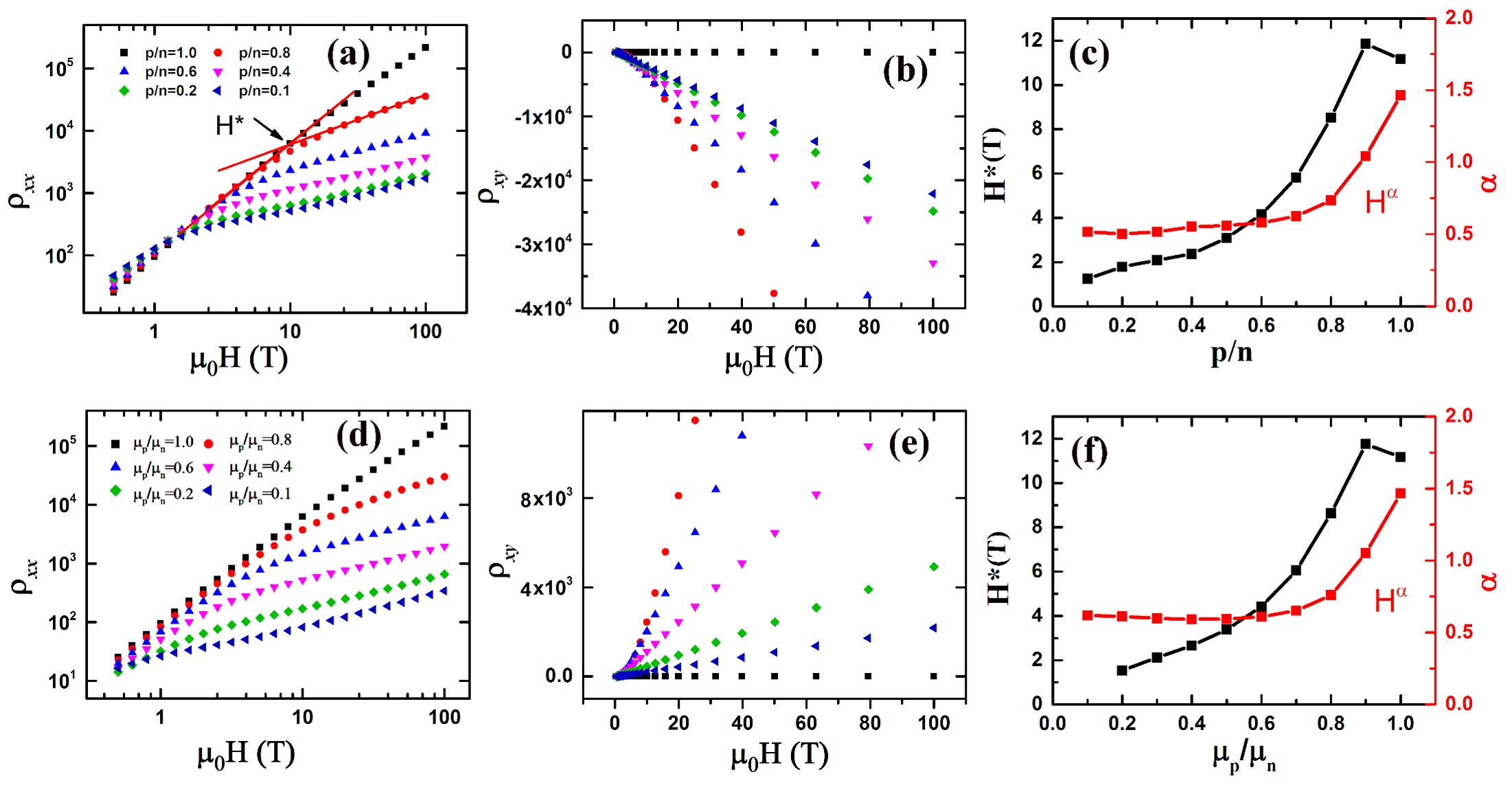}
\end{center}
\caption{The magnetotransport of Weyl semimetal at n greater than p or $\mu_n$ greater than $\mu_p$.(a)-(c) at fixed $\mu_{n,p}=10^5cm^{2}V^{-1}s^{-1}$. (a)The longitudinal resistance $\rho_{xx}$ as a function of magnetic field at different carrier ratios p/n. (b) the corresponding transverse resistance $\rho_{xy}$ as a function of magnetic field. (c)The "turn-on" magnetic field (black line) and the power law at high magnetic field (red line) as a function of the carrier ratio p/n from 0.1 to 1. (d)-(f) at fixed $n=p$ and $\mu_n=10^5cm^{2}V^{-1}s^{-1}$. (d)The longitudinal resistance $\rho_{xx}$ as a function of magnetic field at different mobility ratios $\mu_p/\mu_n$. (e) the corresponding transverse resistance $\rho_{xy}$ as a function of magnetic field. (f)The "turn-on" magnetic field (black line) and the power law at high magnetic field (red line) as a function of the mobility ratio p/n from 0.1 to 1. }
\label{fig:fig3}
\end{figure}

\begin{figure}[t]
\begin{center}\includegraphics[width=\columnwidth]{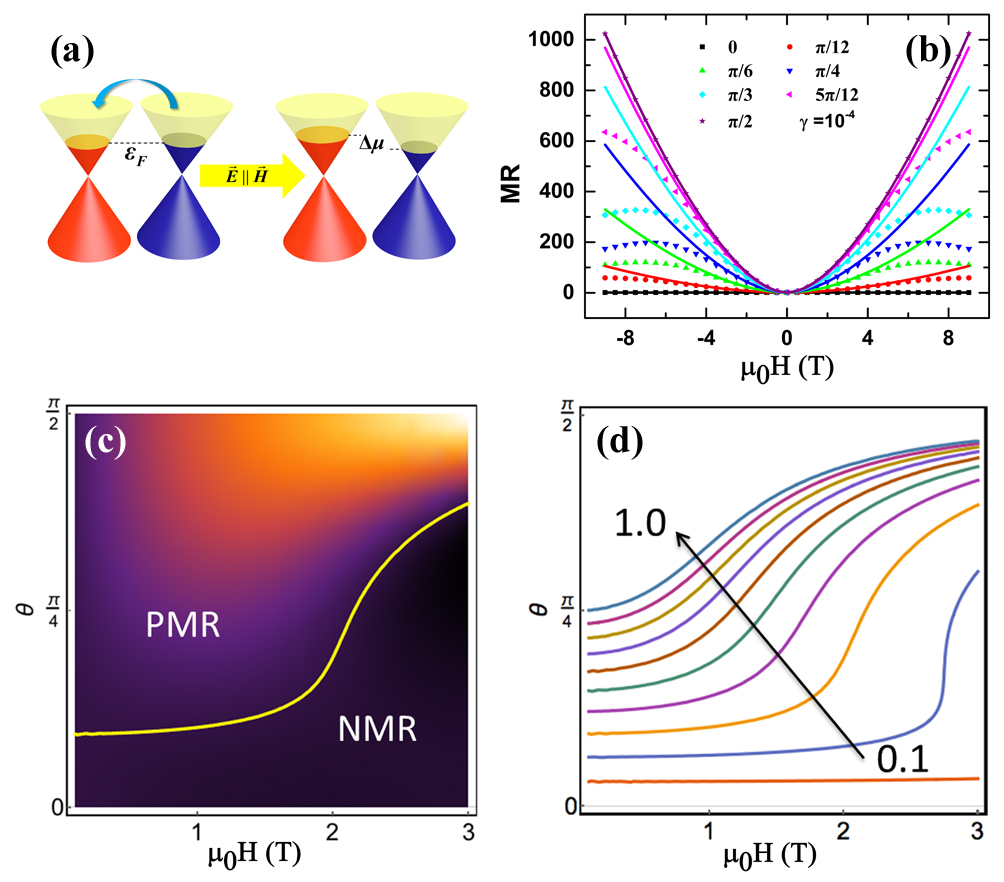}
\end{center}
\caption{The angle-dependence magnetoresistance from positve MR to negative MR. (a)The chiral energy difference induced by chiral anomaly within parellel magnetic field H and electric field E. (b)The MR (points) as a function of magnetic field at different angle from 0 to $\pi/2$ and fixed mobility $\mu_{n,p}=10^5cm^{2}V^{-1}s^{-1}$.The dashed lines show the positive MR without chiral anomaly as a reference. (c) The density plot of MR as a function of magnetic field and angle. The yellow line shows a transition between PMR to NMR at $\delta=0.3$. (d) The contour plot of magnetoresistance transition at different parameter $\delta$. The NMR region enlarges with $\delta$ increasing.}
\label{fig:fig4}
\end{figure}

\end{document}